# Measuring Research Difficulty of Academic Papers: A Case Study in Natural Language Processing

Haochuan Li, Jingyuan Li, Yi Zhao, Heng Zhang, Yukai Yang, Zile Hu, Chengzhi Zhang

*Department of Information Management, Nanjing University of Science and Technology, Nanjing, China* 210094

**Abstract**: With the rapid growth of the number of academic papers, systematically evaluating the difficulty of research and its relationship to academic impact offers important significance for research topic selection and resource allocation. However, current studies lack quantitative assessments of research difficulty and its correlation with academic impact. This paper proposes a comprehensive evaluation system for research difficulty, incorporating factors such as academic collaboration, content, and references. Taking the field of Natural Language Processing (NLP) as a case study, we extract both internal and external features from academic papers, compute multiple research difficulty indicators. We assign their weights using the entropy weight method and perform a weighted sum to obtain the research difficulty score of academic papers. This paper uses the citation frequency of academic papers to measure academic impact. To validate our approach, NLP experts assessed the difficulty of a sample of papers, and correlation analyses confirmed the reliability of our measurement. Empirical results reveal that in NLP, factors such as the number of pages, reference count, and participation of high-level institutions are significantly associated with academic impact. Moreover, we identify an inverted U-shaped relationship between research difficulty and academic impact. It suggests that moderately difficult research tends to achieve greater academic impact.

## 1. Introduction

Difficulty is the level of challenge inherent in tasks, whether in work, technology, or other areas. As the German thinker Goethe stated, "The closer the goal, the greater the difficulty." Generally speaking, it is relatively easy to make progress in the early stages of scientific research. However, as research advances, standards for optimal outcomes continue to rise. It makes it difficult to get huge breakthroughs in science and technology. During the scientific and technological development, absolute difficulty consistently grows, while relative difficulty gradually diminishes. This results in higher levels of achievement and shorter intervals between milestones. Academic papers, a medium and tool for scientific communication, aim at disseminating culture and knowledge, as well as exerting influence and impact on others. The research difficulty of academic papers refers to the challenges researchers encountered in research process. Since 1980, the number of papers published in scientific literature has grown exponentially at an approximate annual rate of 3% (Bornmann & Mutz, 2015). Measuring the research difficulty of academic papers enables a systematic assessment of the challenges researchers face in academic activities.

Quantitative assessments of research difficulty in academic papers are relatively rare. Existing studies tend to focus on evaluating factors such as paper innovativeness (Wang et al., 2024), impact (Ha, 2022), and readability (Lei & Yan, 2016), with limited examination of the relationship between research difficulty and academic impact. Most current studies focus on how different factors relate to citations. For example, they study the relationship between the number of authors and citations, the number of references and citations, and the length of the paper and citations.(Borsuk et al., 2009; Ahmed et al., 2016). Studying how research difficulty affects academic impact and looking at how knowledge spreads from papers of different difficulty levels can help researchers understand the influence of difficulty on impact. This study shows research characteristics in subfields. It helps with choosing topics and also supports better use of research resources and making good research strategies and policies.

This study uses the field of Natural Language Processing (NLP) as a case study to construct an evaluation system for assessing the research difficulty of academic papers. Factors such as collaboration, figures and tables, knowledge entities, and references are incorporated into the assessment. Features are extracted from NLP papers. Research difficulty indicators are calculated across different dimensions. We use the entropy weight method to determine the weights of these indicators. Finally, the weights are summed to produce a research difficulty score. Empirical analysis confirms the reliability of this method. Additionally, in the NLP field, findings show that page count, reference count, and participation by high-ranking institutions are positively correlated with academic impact. And there is also an inverted U-shaped relationship between research difficulty and academic impact.

## 2. Related Work

This section provides an overview of related work from three perspectives: difficulty measuring of task-specific, difficulty measuring of textual content and factors influencing academic impact of papers

### 2.1 Difficulty measuring of task-specific

Research on task difficulty spans multiple fields, each adopting distinct evaluation methods for difficulty. In the field of information retrieval, numerous studies focus on task difficulty. Pre-task difficulty measurement is often determined by researchers or evaluated by users. For instance, Bell & Ruthven (2004) studied users' perceptions of task difficulty in web search environments, finding that task difficulty and complexity were closely related concepts. Researchers can also simulate retrieval tasks and assess difficulty based on how challenging it is to match task descriptions with target retrieval results. For example, Wacholder & Liu (2006) preset task difficulty in the ESBI information retrieval system by evaluating whether terms in the problem description could directly serve as query terms to measure task difficulty. Another approach is to assess task difficulty based on the relevance of the target retrieval results. Bailey et al. (2009) chose 50 TREC topic tasks. They calculated the nDCG of the top 50 retrieval results to evaluate relevance, finding that higher average nDCG means lower task difficulty. Post-task difficulty measurement has two categories. One uses interviews and surveys to get user feedback directly (Gwizdka & Lopatovska, 2009; Liu et al., 2012). The other evaluates difficulty based on users' actual search processes and outcomes. For example, Aula et al. (2010) analyzed the success rate of retrieval results when users encountered search challenges, finding that the fewer users who succeeded, the more difficult the task. Mack et al. (2004) argued that measuring task difficulty should occur after users complete retrieval tasks with clear answers.

In other fields, research on task difficulty involves measuring the difficulty of tasks such as speaking, listening, and software development. For example, Fulcher and Reiter (2003) defined a new method of speaking task difficulty, considering the interaction between practical task characteristics and the cultural background of the speaker's first language. They explored the impact of these variables on test scores, as well as the generalizability of score meanings and the relevance of structure definition in speaking tests. In studies on listening task difficulty, Brindley and Slatyer (2002) analyzed test scores. They found that speaking speed and question format have a big effect on task and question difficulty. Regarding software development tasks, Fritz et al. (2014) used data from psychophysiological sensors to categorize the difficulty of code comprehension tasks. Their work established a foundation for developing feasible and reliable task difficulty measurement standards, which are expected to support next-generation programming tools.

In general, various fields have accumulated rich research results in the evaluation of task difficulty. Research on the measurement of specific task difficulty provides a reference for the selection of research difficulty indicators in this academic paper.

## 2.2 Difficulty measuring of textual content

Evaluating and analyzing the difficulty of textual content is a key aspect of readability analysis, which dates back to the 1920s.(Vogel & Washburne, 1928). Automatic readability analysis is a quantitative and automatic assessment and analysis of the difficulty of text content. Its analysis methods can be subdivided into formula method, classification method, and ranking method.

The formula method predicts text difficulty by establishing linear equations and using language features closely related to text difficulty as variables. Commonly used features include word length and sentence length. For example, in the 1920s, Vogel and Washburne (1928) were the first to use regression equations with multiple text features in readability formulas. This had a big influence on later readability research. Kincaid (1975) developed a readability analysis system centered around readability formulas to evaluate and grade national policy documents. The formula method is objective and straightforward, providing a quantified approach to assessing reading difficulty and laying the foundation for readability research. However, these formulas cannot account for all variables affecting readability, which has raised questions about their validity.

The classification method treats text difficulty prediction as a classification task, building classification models by learning distinguishing features from texts of different difficulty levels. After entering new text without labels, the classification model estimates the difficulty level of the text based on the results it has learned. For example, Si and Callan (2001). Using a unigram string membership model in text readability analysis. Petersen and Ostendorf (2009) used support vector machines for readability analysis. They used text features from n-gram models, as well as lexical and syntactic features, during the training process. Previous research results have shown that classification methods often yield more accurate readability scores than formula methods.

The ranking method constructs a comparator or manually labels the relative difficulty of each pair of texts, then sorts the texts to obtain a set of texts arranged by difficulty. However, this method does not yield specific difficulty values or levels. Tanaka-Ishii et al. (2010) used a ranking-based method to measure text difficulty without labeled data. They compared texts using greater than, less than, and equal to. Schumacher et al. (2016) got pairwise comparison data manually and used a scoring ranking algorithm to rank texts. This method matches how people compare text difficulty.

In summary, each of these three methods for assessing textual content difficulty has its strengths and limitations. The choice of method depends on the specific application and research objectives.

## 2.3 Factors influencing academic impact of papers

Academic impact helps scholars quickly and comprehensively assess the value of academic papers. Therefore, to meet the demands of modern developments, researching the factors related to the impact of academic papers is essential. This paper reviews representative studies in recent years that examine factors related to academic impact (Table 1). As shown, the number of authors in various fields has a significantly positive influence on academic impact (Annalingam et al., 2014; Robson & Mousquès, 2014). Additionally, factors such as paper length (Tahamtan et al., 2016), journal quality (Tahamtan et al., 2016), and algorithm frameworks used (Kang et al., 2023) also significantly impact a paper's influence. However, linguistic complexity does not have a significant relationship with academic impact.(Lu et al., 2019).

In summary, existing research explores various factors affecting academic impact from multiple perspectives. There's not much research on the relationship between research difficulty and academic impact. This paper aims to create an evaluation system to measure research difficulty in academic papers. We also focus on this relationship in NLP. Furthermore, we will examine differences in the relationship between research difficulty and academic impact across different NLP tasks.

**Table 1. Literature Review of Factors Related to Academic Influence**

| Author | Research Content | Research Finding |
|---|---|---|
| Annalingam et al. (2014) | This study, based on the SciVerse Scopus database, analyzed the relationship between journal ranking, number of authors, and the number of citations | Journal rankings, number of authors, overseas research, systematic review(meta-analysis), and regional or international cooperation have all significantly increased citations. |
| Robson and Mousques(2014) | This study analyzed whether it is possible to predict the number of citations obtained by each paper and explore the factors that affect the number of citations under the condition of uncertain paper quality | The papers with longer length and review type are cited more, the papers containing differential or integral equations are cited less, and the increase of citations is related to the increase of the number of authors. |
| Tahamtan et al. (2016) | The study included the influence of paper quality, research field, paper title, abstract and keyword characteristics, the number and quality of references, paper length, early citations, downloads, journal influence and other factors on the number of citations of academic papers. | There are many factors that affect the number of citations of academic papers, among which the quality of papers, research field, length of papers, number of citations obtained in the early stage of publication, influence of journals and other factors have significant influence. There are interactions between different factors, which jointly affect the academic influence of the paper. |
| Lu et al. (2019) | The language complexity variable was selected to measure the complexity of the paper and analyze its relationship with citation. | There is no actual significant relationship between language complexity and reference layer. |
| Kang et al. (2023) | The study evaluated how the GitHub repository affects the citation rate of machine learning papers and the impact of model trends on the citation rate of papers | Compared with papers without such repositories, the monthly citation rate of machine learning papers increased significantly after the creation of their first GitHub repository, and the trend of using different machine learning frameworks over time affected the monthly citation rate. |

## 3. Methodology

This paper measures the research difficulty of academic papers in the field of NLP and examines the relationship between research difficulty and academic impact. This section primarily introduces the key components, including the construction of the research difficulty evaluation system for academic papers, the acquisition of research difficulty indicators, the calculation of research difficulty scores, and the analysis of the relationship between research difficulty and academic impact.

### 3.1 Construction of research difficulty evaluation system for academic papers

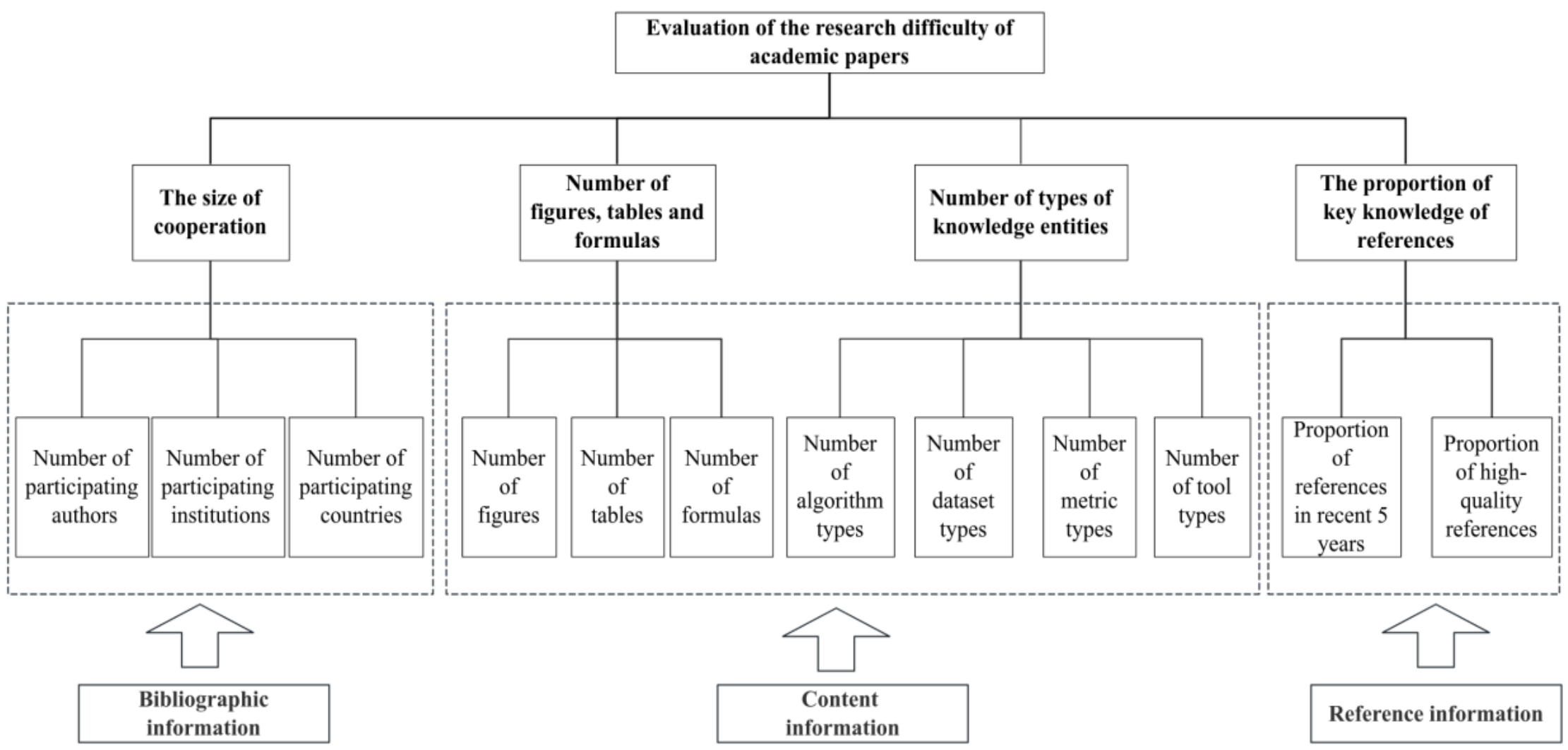

**Fig.1. Research difficulty evaluation system of academic papers**

Wang et al. (2024) suggested that the quality evaluation of academic papers could be structured around coarse-grained features (e.g., number of images, tables, formulas, references) and knowledge entities (e.g., macro entities, micro entities). Therefore, we define the research difficulty of academic papers as the level of various challenges and obstacles encountered during the research process. In other words, the difficulties encountered in terms of technology, theory, methodology, and social factors, as well as the efforts expended in overcoming them, can be used to measure the difficulty of academic research issues. Based on these considerations, we developed a preliminary framework for evaluating the research difficulty of academic papers. We then consulted three experts in the field of Natural Language Processing and adjusted the evaluation indicators according to their feedback. The final evaluation system for academic paper research difficulty is shown in Figure 1. This system considers three aspects: bibliographic information, content, and references of academic papers.

#### (1) Indicators of research difficulty based on academic paper bibliographic Information

The scale of collaboration on an academic paper reflects the effort researchers invest in addressing social aspects of a research problem. A paper’s bibliographic information generally includes title, authors, keywords, author affiliations, and abstract. Using the authorship information from the bibliographic data, we can determine the scale of collaboration for a paper. We identify three secondary indicators under the primary indicator of collaboration scale: the number of participating authors, number of participating institutions, and number of participating countries, which represent the collaborative scope across scholars, institutions, and nations involved in the research.

#### (2) Indicators of research difficulty based on academic paper content

The content of an academic paper typically reveals specific details of the research methods used. By examining the full text, we can assess the challenges and difficulties encountered during the research implementation. We divide the research difficulty indicators based on content into two primary indicators: the number of figures, tables, and formulas, and the diversity of knowledge entities.

Figures, tables, and formulas illustrate details of research methods, experimental data, or theoretical models. Figures and tables can evaluate the scientific level of a scientific field to some extent (Ma & Li, 2022). Formulas reflect the mathematical methods involved, representing the computational or reasoning processes in research. Therefore, we define three secondary indicators: the number of figures, the number of tables, and the number of formulas.

For the diversity of knowledge entities, we consider the models, algorithms, datasets, tools, and evaluation metrics used in the research. The main indicator includes four secondary indicators: the number of algorithm types, dataset types, tool types, and metric types.

**(3) Indicators of research difficulty based on academic paper references**

References in academic papers are essential for providing theoretical foundations, evaluating prior research, ensuring academic integrity, and fostering scholarly exchange and collaboration. References partly reflect the process of building theoretical knowledge within a paper. Based on characteristics of NLP, we define two secondary indicators under the primary indicator of references: the proportion of references from the past five years and the proportion of high-quality references. The proportion of references from the past five years refers to the proportion of references from the five years prior to the publication of the cited paper. This indicator reflects the extent to which academic papers explore cutting-edge topics. The proportion of high-quality references reflects the share of citations from high-impact publications, indicating the quality and credibility of the references used.

## 3.2 Acquisition of research difficulty indicator values

This section first introduces the process of obtaining academic papers in NLP field and then describes the procedure for extracting research difficulty indicators.

### 3.2.1 Obtaining full texts of academic papers in the NLP field

We measure research difficulty using NLP as a case study. The research data are selected from the main conference papers of three major international conferences in the NLP field: ACL (Annual Meeting of the Association for Computational Linguistics), EMNLP (Conference on Empirical Methods in Natural Language Processing), and NAACL (Annual Conference of the North American Chapter of the Association for Computational Linguistics). ACL Anthology [1] provides downloadable full-text PDFs of all papers presented at ACL, EMNLP, and NAACL conferences. We downloaded the full-text PDFs of main conference papers from these three conferences held between 2000 and 2021, totaling 13,772 papers. Additionally, we scraped metadata for each paper, including title, authors, abstract, and year of publication. The fields of the final dataset are shown in Table 2.

[1] https://aclanthology.org/

Table 2. Bibliographic Information of Academic Papers in the NLP Field

| Type | Example |
|---|---|
| Paper ID | P15-1139 |
| Conference | ACL |
| Time | 2015 |
| Title | Tea Party in the House: A Hierarchical Ideal Point Topic Model and Its Application to Republican Legislators in the 112th Congress |
| Author | Nguyen Viet-An, Boyd-Graber Jordan, Resnik Philip and Miler Kristina |
| Abstract | We introduce the Hierarchical Ideal Point Topic Model… (Note: Due to space limitation, the remaining part is omitted.) |

### 3.2.2 Acquisition of research difficulty indicator values for academic papers in the NLP field

#### (1) Collaboration scale

The bibliographic information of academic papers typically includes details such as author names and their institutions. Using a combination of web scraping and manual annotation, we obtained the names of authors, their institutions, and the countries of those institutions for each paper presented at the three major conferences. Based on this information, we calculated the number of participating authors, institutions, and countries for each paper in these NLP conferences.

To determine the number of institutions collaborating on each paper, we first extracted the names of all institutions from the bibliographic information. Then, we counted distinct institution names associated with each paper ID to obtain the number of participating institutions per paper. Next, we matched these institutions with their respective countries using the Global Research Identifier Database (GRID) [2]. For institutions without matched country information, we conducted manual searches to supplement this data. Finally, we counted the countries involved for each paper to determine the number of participating countries.

#### (2) Number of figures, tables, and formulas

After downloading the PDF-format papers from the three major NLP conferences, we used the open-source tool pdf2htmlEX[3] to convert them into HTML format. Based on this, we developed rules to extract the number of figures, tables, and formulas for each paper. The specific process is as follows.

- **Extraction of number of figures and tables**

We first sampled 50 papers to observe keywords associated with figures and tables, creating an initial list of keywords. Using this list along with section titles, we identified the locations and sequence numbers of figures and tables within each paper. The total number of figures and tables was then determined by the highest sequence number following each keyword. And we corrected by checking the continuity of the serial numbers and the position of figures/tables. We randomly selected 100 papers to manually verify the accuracy of the extraction, achieving an accuracy rate of 96% for figures and 94% for tables.

- **Extraction of number of formulas**

[2] https://www.grid.ac/

[3] https://pdf2htmlex.github.io/pdf2htmlEX/

We randomly sampled 50 papers to examine characteristics of formulas within academic papers. As a result, we found that formulas are typically accompanied by unique identifiers, line breaks, indentation, special symbols, or specific keywords. Based on these features, we identified and counted the number of formulas in each paper. To verify accuracy, we randomly selected 100 papers for manual review, and results indicated an accuracy rate of 91% for formula extraction.

**(3) Number of knowledge entity types**

Currently, pre-trained language models are widely used in the field of information extraction, with popular models including BERT and SciBERT. BERT (Bidirectional Encoder Representations from Transformers), developed by Google, is a pre-trained language model designed to improve the performance of natural language processing tasks through bidirectional encoder representations (Devlin, et al., 2018). SciBERT, proposed by Beltagy et al. (2019), is a BERT-based variant pre-trained specifically on scientific literature. Additionally, Wei et al. (2019) introduced the Cascade Binary Tagging Framework, enhancing the model's ability to recognize entity boundaries. In this paper, we construct an entity recognition model based on SciBERT and the Cascade Binary Tagging Framework. It optimizes the framework by adding a sequence to encode entity types, thus adapting it for multi-entity recognition tasks. This model is represented as SciBERT+BiLSTM(cascade). Four baseline models—BERT, SciBERT, BERT+CRF, and SciBERT+CRF—are also constructed, with the performance evaluation results of each model on the entity annotation dataset shown in Table 3.

**Table 3. Evaluation results of entity extraction model**

| Model | Precision (%) | Recall (%) | $F_1$ (%) |
|---|---|---|---|
| BERT | 87.87 | 82.14 | 84.91 |
| BERT+CRF | 85.56 | 80.88 | 83.15 |
| SciBERT | 84.77 | **86.55** | 85.65 |
| SciBERT+CRF | 87.34 | 84.03 | 85.65 |
| SciBERT+BiLSTM(cascade) | **88.16** | 84.45 | **86.27** |

As shown in Table 3, the SciBERT+BiLSTM(cascade) model achieves an $F_1$ score of 86.27%, outperforming the other four baseline models. Therefore, this paper utilizes the SciBERT+BiLSTM(cascade) model to extract four types of fine-grained knowledge entities—algorithms, datasets, tools, and metrics—from the full text of unannotated papers from the three major NLP conferences. Subsequently, high-frequency entity abbreviations are converted to their full names. Based on the method of Zhang et al. (2024), entity similarity is calculated based on morphological characteristics, and clustering methods are used to normalize the entities.

After entity normalization, we deduplicated the extracted knowledge entity data to obtain a unique set of knowledge entities for each paper. Finally, we separately count the number of each type of knowledge entity in each paper.

**(4) Proportion of key knowledge in references**

We use the GROBID[4] tool to extract reference lists from academic papers in NLP and further parses these references to extract details such as author, year, and journal or conference name. Under the primary indicator of the proportion of key knowledge in references, there are two secondary indicators: the proportion of references from the past five years and the proportion of high-quality references. The calculation process for each is as follows.

[4] https://github.com/kermitt2/grobid

- **Proportion of references from the past five years**

Price referred to documents published within the last five years as "current" literature, while documents older than five years are considered "archival" literature. "Current" literature reflects frontier research and the latest trends in the field. We use the proportion of references from the past five years as one of the secondary indicators to assess the research difficulty of academic papers. For example, if Paper A was published in 2010 and cites a total of 20 references, of which 15 were published within the last five years, the proportion of recent references for Paper A would be 15/20*100%=75%.

- **Proportion of high-quality references**

Citing high-quality references is key to ensuring the reliability and academic value of research. The China Computer Federation (CCF) recommended list of international conferences and journals[5] serves as a recognized list of important publications in computer science. This study uses the 2022 CCF recommended list to define high-quality references. If a cited publication appears on the CCF list, it is classified as high quality. For example, if Paper B cites 30 references, of which 15 are from publications on the CCF recommended list, then the proportion of high-quality references for Paper B would be 15/30*100%=50%.

### 3.3 Calculation of research difficulty score for academic papers

Due to differences in dimensions and units among the indicators, it is necessary to standardize the results for each indicator. This process maps data with different dimensions to a common comparable range, allowing for comparisons on the same scale. We apply the min-max normalization method to standardize the indicator results, as shown in Equation (1).

$$y_{ij} = \frac{x_{ij} - Min_j}{Max_j - Min_j} \tag{1}$$

Where, $y_{ij}$ represents the normalized value, $x_{ij}$ is the value of paper *i* on indicator $j$, $Max_j$ is the maximum value of indicator $j$, and $Min_j$ is the minimum value of indicator $j$. Since the min-max normalization method is sensitive to outliers, this study excludes extreme high and low values based on the data distribution characteristics and treats these outliers as the upper and lower limits during normalization—assigning a value of 1 to larger outliers and 0 to smaller outliers.

After normalizing the 12 secondary indicator values shown in Figure 1, we used the entropy weight method to calculate the weight of each indicator (Ke et al., 2009), as outlined in Equations (2) to (6).

$$\rho_{ij} = \frac{y_{ij}}{\sum_{i=1}^{n} y_{ij}} \tag{2}$$

Where, $\rho_{ij}$ represents the proportion of the *i*-th sample value in indicator $j$, and $y_{ij}$ denotes the normalized value of paper *i* for indicator $j$.

$$e_j = -\frac{1}{\ln(n)} * \sum_{i=1}^{n} \rho_{ij} * \ln(\rho_{ij}) \tag{3}$$

Where, *n* represents the total number of samples, and $e_j$ denotes the entropy value of indicator $j$.

$$d_j = 1 - e_j \tag{4}$$

[5] https://www.ccf.org.cn/Academic_Evaluation/By_category/

Where, $d_j$ represents the divergence coefficient of indicator $j$.

$$w_j = \frac{d_j}{\sum_{j=1}^{m} d_j} \tag{5}$$

Where, $m$ is the total number of indicators, which is 12 in this study, and $w_j$ is the weight of indicator $j$.

Finally, by performing a weighted summation of all indicators, we obtained the research difficulty score for each paper in the three major NLP conferences, as shown in Equation (6).

$$Diffculty_i = \sum_{j=1}^{m} y_{ij} * w_j \tag{6}$$

Where, $Diffculty_i$ represents the research difficulty score of paper $i$, $y_{ij}$ is the score of indicator $j$ for paper $i$, and $w_j$ is the weight of indicator $j$.

The final calculated weights for the 12 indicators in this study are shown in Table 4. Indicators such as the number of participating countries, the number of participating institutions, and the number of formulas have relatively high weights, all exceeding 0.1. It indicates their importance in measuring the research difficulty of academic papers. A higher number of participating countries and institutions generally suggests broader international cooperation and interdisciplinary collaboration. A larger number of formulas reflects the theoretical depth and mathematical complexity of the research content. In contrast, reference-related indicators, such as the proportion of references from the past five years and the proportion of high-quality references, have lower weights. It suggests that while recent and high-quality references are critical for supporting and substantiating research conclusions, they are less important for evaluating the research difficulty of academic papers.

**Table 4. Weight of academic paper research difficulty evaluation index**

| Metrics | Weight | Metrics | Weight |
|---|---|---|---|
| Number of participating authors | 0.0457 | Number of dataset types | 0.0469 |
| Number of participating institutions | 0.1702 | Number of algorithm types | 0.0263 |
| Number of participating countries | 0.3184 | Number of tool types | 0.0973 |
| Number of figures | 0.0640 | Number of metrics types | 0.0400 |
| Number of tables | 0.0451 | Proportion of references in recent 5 years | 0.0135 |
| Number of formulas | 0.1193 | Proportion of high-quality references | 0.0133 |

### 3.4 Analysis of relationship between research difficulty and academic impact of academic papers

After measuring the research difficulty of academic papers, we use regression analysis to explore the relationship between research difficulty and academic impact. The research difficulty score of academic papers is used as the independent variable, while academic impact is the dependent variable. This section provides a detailed description of the measurement of academic impact and the control variables used in the regression analysis.

**(1) Measurement of academic impact of academic papers**

Citation frequency is one of the most common methods for measuring the academic impact of papers. Previous studies have shown that citation frequency typically rises rapidly in the first few years after publication (Bornmann & Williams, 2013). For most papers, citations within the first five years are highly correlated with total citation counts (Wang, 2013). Therefore, citations within the first five years after publication can largely represent a paper's academic impact. In this study, we use the citation frequency within five years after publication to measure the academic impact of papers presented at the three major NLP conferences.

We selected a total of 8,447 papers published between 2000 and 2018 at these conferences. The citation frequency within five years of publication was retrieved from Google Scholar[6], with data collected in April 2023. For papers published less than five years ago, the total citation count is used as the measure of academic impact.

Due to the significant variation in academic impact among papers in the NLP field, and the presence of papers with zero citations, we apply a log transformation (adding one) to citation frequency to better fit the regression model. This transformed value is used as the dependent variable in the regression model.

**(2) Control variables**

The control variables used in this study include six factors: paper length, number of references, collaboration type, involvement of high-ranking institutions, conference category, and publication year[7].

Increasing paper length allows for more content, which can enhance citation frequency to some extent (Perneger, 2015). To obtain paper length, we utilize the readability of the HTML format and developed rules to calculate page numbers. Sampling tests showed an accuracy rate of 100%.

The number of references reflects a paper's ability to absorb and utilize knowledge. In certain fields, highly cited papers often contain more references (Smith et al., 2014).

Collaboration lets people share resources and use each other's strengths. International collaborations bring together the advantages of different countries. Previous research indicates that institutional collaboration can influence citation frequency to varying degrees, and international collaborations may positively impact citation counts (Smith et al., 2014; Puuska et al., 2014). Therefore, we included collaboration type as a control variable, categorized as single institution research (CT0), intra-institutional collaboration (CT1), inter-institutional collaboration within the same country (CT2), and international collaboration (CT3).

Papers involving high-ranking institutions may receive higher citation counts (Garner et al., 2014). Thus, we included the presence or absence of high-ranking institution involvement as a control variable. High-ranking institutions are defined as the top 500 universities in CS Rankings[8] and the top 500 companies in the "Largest Companies by Market Cap[9]." CS Rankings is an authoritative ranking site for computer science, and we obtained the top 500 data from the 2022 CS Rankings using web scraping. "Largest Companies by Market Cap" ranks companies by market value, and we collected data on the top 500 companies as of August 4, 2022. After compiling high-ranking institution data, we coded the presence or absence of such institutions as 1 and 0,

---

[6] https://scholar.google.com/

[7] So et al. (2015) found a significant correlation between the number of authors, paper length, and citation frequency; multiple authors can pool their strengths to conduct research with greater academic value. Given that the research difficulty evaluation system in this study already includes the "number of participating authors" as an indicator, it is excluded as a control variable in the regression analysis.

[8] https://csrankings.org/

[9] https://companiesmarketcap.com/

respectively.

Garner et al. (2014) suggested that the impact of the publication venue can influence citation frequency to some extent. Although we control for citation time span, publication year may still affect citation frequency through other factors. Therefore, we included conference category (where EMNLP and NAACL are relative to ACL) and publication year as control variables.

# 4. Result Analysis

This study conducts an empirical analysis of the research difficulty evaluation system for academic papers, using NLP conference papers as an example. First, we invited domain experts to assess the research difficulty results. Next, the measurement results of research difficulty were examined from two perspectives: the overall NLP field and different NLP tasks. Finally, a regression model was used to analyze the relationship between research difficulty and academic impact.

## 4.1 Evaluation of research difficulty results for academic papers

To verify the reliability of the research difficulty measurement results for academic papers in the NLP field, we invited NLP experts to rate the research difficulty of papers in this domain. A 5-point scale was used to assess research difficulty, with "1," "2," "3," "4," and "5" indicating "very low research difficulty," "low research difficulty," "moderate research difficulty," "high research difficulty," and "very high research difficulty," respectively. We selected eight NLP task categories with the highest paper counts in the three major NLP conferences. These categories are machine translation, semantics, information extraction, analysis, machine learning, representation learning and text representation, resources and evaluation, and sentiment analysis and opinion mining. For each of these tasks, five papers were randomly selected, totaling 40 papers. Five NLP experts familiar with these task areas were invited to evaluate the research difficulty of these papers. Experts assessed each paper's research difficulty based on the abstract or full text.

Based on the experts' evaluations, the average difficulty rating for each paper was calculated. Results showed that among the 27 papers with difficulty scores above 0.3, only two papers had an average expert rating below 3. Additionally, of the 10 papers with an average expert rating of 4 or above, only one had a difficulty score below 0.3, and two had scores below 0.35.

**Table 5. Correlation Between Research Difficulty Scores and Expert Evaluations**

| | **Attribute** | **Experts Score** |
|---|---|---|
| **Academic paper research difficulty score** | Spearman Correlation Coefficient | 0.459** |
| | Sig.（Bilaterally） | 0.000 |
| | N | 40 |

**Note**: ** indicates that the correlation is significant at a 0.01 confidence level (two-tailed).

Then, we conducted a Spearman correlation analysis between the average expert rating of research difficulty for each paper and the research difficulty scores calculated. The results are shown in Table 5. The analysis reveals a significant positive correlation between the calculated research difficulty scores and the expert ratings at the 0.01 level ($p < 0.01$), with a Spearman correlation coefficient of 0.459. This shows that the method in this study is very reliable. It proves the calculation method is accurate and the difficulty scoring indicators can truly reflect the research difficulty of academic papers.

## 4.2 Analysis of research difficulty results for academic papers

This section analyzes the research difficulty results of academic papers in the NLP field from three perspectives: descriptive statistics of research difficulty, a comparison of research difficulty across different NLP task categories, and the evolution of research difficulty in the NLP field.

**(1) Descriptive statistical analysis of research difficulty in academic papers within the NLP Field**

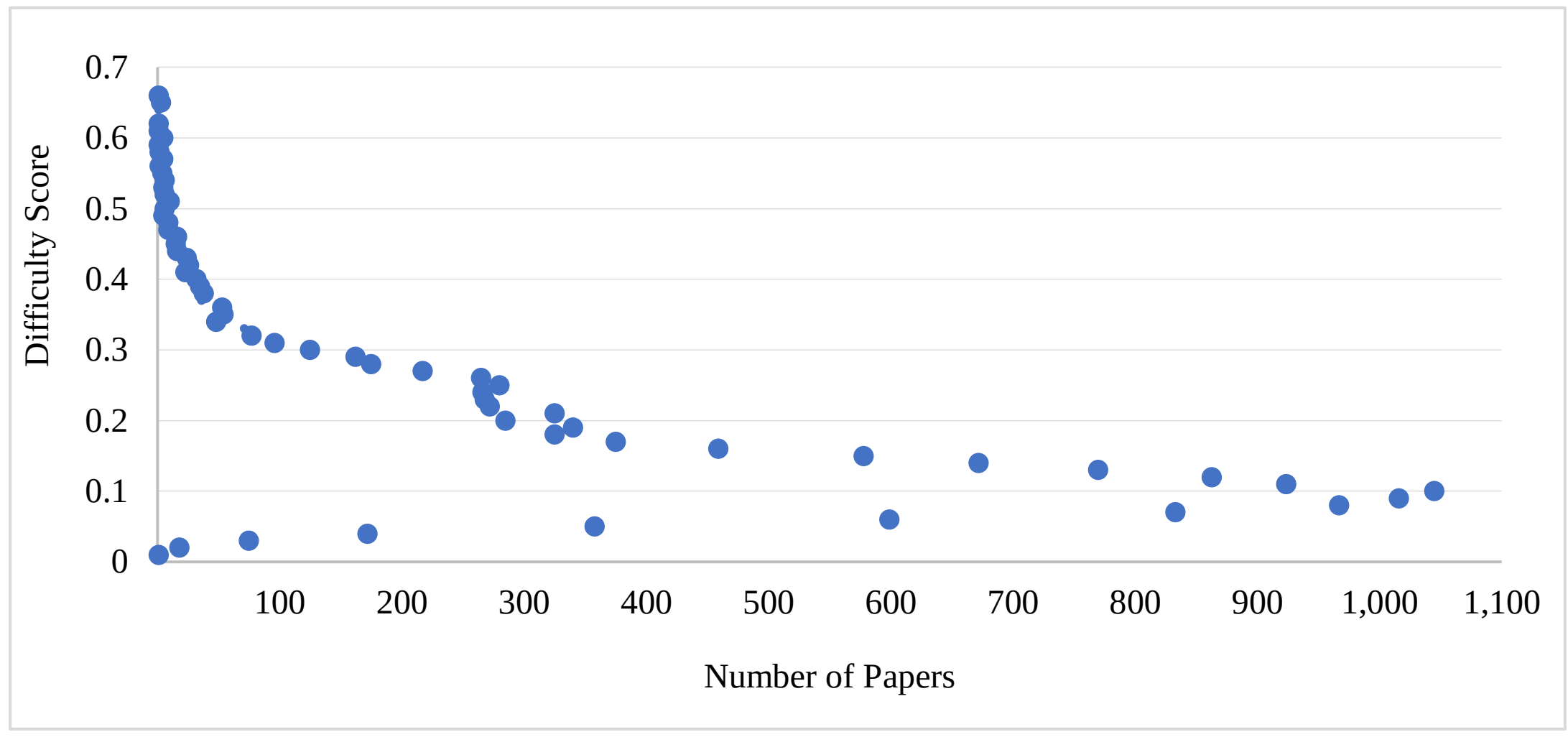

**Fig.2. Scatter plot of research difficulty distribution of academic papers in NLP field**

The distribution of research difficulty scores for NLP academic papers from 2000 to 2021 is shown in Figure 2. Most papers' research difficulty scores below 0.2. This indicates that in the field of NLP, most papers are relatively easy to research, while only a few are difficult to research. This reflects an uneven distribution of research difficulty in NLP academic papers. It might be because NLP technology is growing fast. Basic tasks are well-developed and easy to study, so they are less difficult. New and creative tasks are harder and not studied as much. This makes the difficulty uneven overall.

This study lists the top ten papers in the NLP field by research difficulty score in Table 6. The results show that seven out of the top ten papers were published after 2020. This showes that research difficulty in the NLP field has increased over time. Moreover, although these papers exhibit higher research difficulty, their academic impact has not reached the expected levels, showing a trend inconsistent with their ranking. This suggests that research difficulty and academic impact may not be positively correlated; that is, papers with greater research difficulty do not necessarily have higher academic impact. This could be because high-difficulty research is more challenging to be understood and widely accepted.

**Table 6. Top-10 papers in research difficulty score**

| Title | Year | Difficulty score | Citation within 5 years |
|---|---|---|---|
| From Masked Language Modeling to Translation: Non-English Auxiliary Tasks Improve Zero-shot Spoken Language Understanding | 2021 | 0.6643 | 6 |
| A Large-Scale Study of Machine Translation in Turkic Languages | 2021 | 0.6530 | 5 |
| KILT: a Benchmark for Knowledge Intensive Language Tasks | 2021 | 0.6482 | 188 |
| ParGramBank: The ParGram Parallel Treebank | 2013 | 0.6477 | 32 |
| The CoNLL 2007 Shared Task on Dependency Parsing | 2007 | 0.6385 | 290 |
| How (Non-)Optimal is the Lexicon? | 2021 | 0.6201 | 10 |
| ParaCrawl: Web-Scale Acquisition of Parallel Corpora | 2020 | 0.6137 | 121 |
| Intrinsic Probing through Dimension Selection | 2020 | 0.6040 | 31 |
| Lightweight and Efficient Neural Natural Language Processing with Quaternion Networks | 2019 | 0.6027 | 59 |
| Revisiting the Uniform Information Density Hypothesis | 2021 | 0.6019 | 19 |

**(2) Comparative analysis of research difficulty across different task categories in the NLP Field**

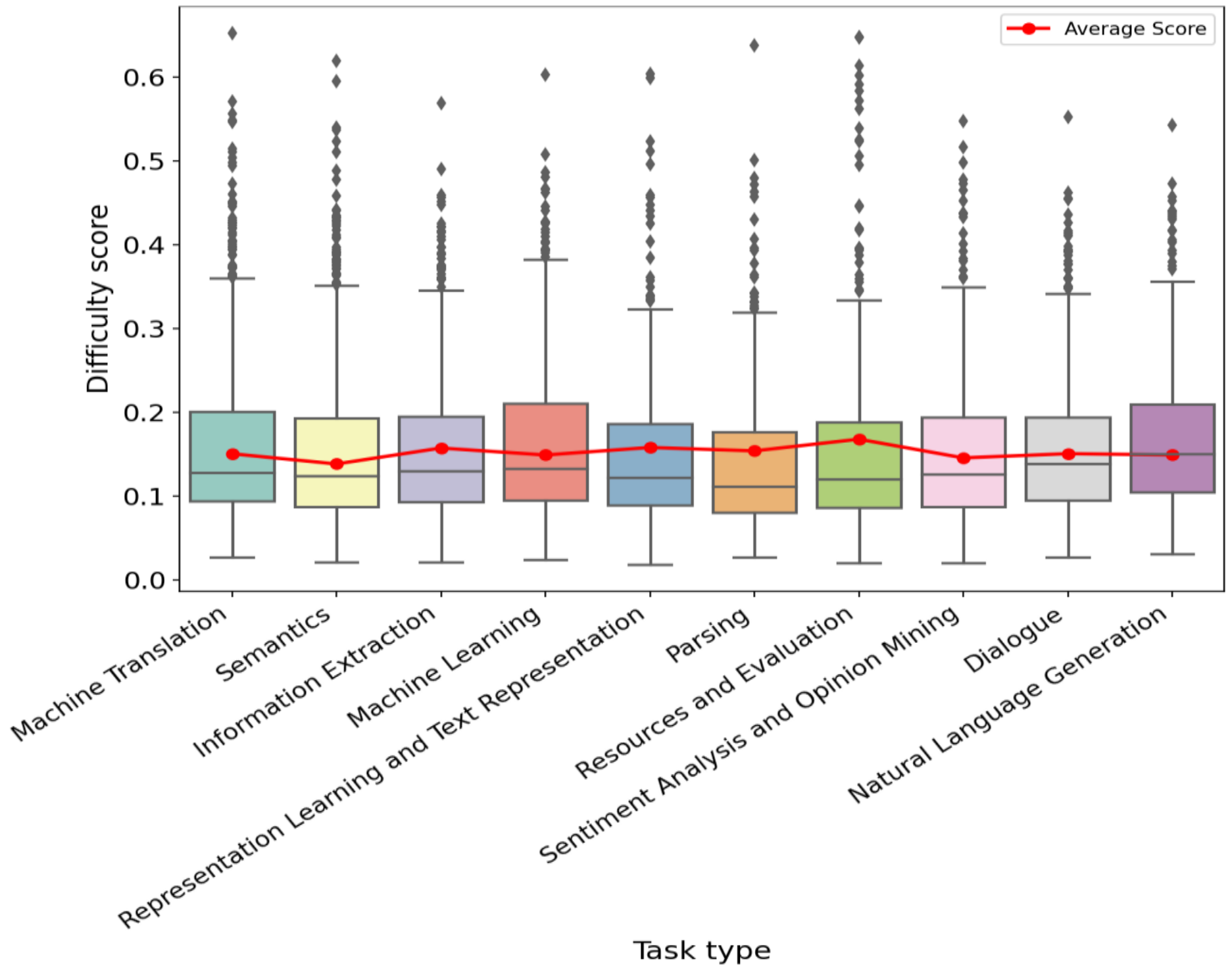

**Fig.3. Different task categories papers research difficulty score box plot**

According to the classification system standards in The Oxford Handbook of Computational Linguistics (Sampson, 2009) and Wang & Zhang (2018), NLP tasks can be divided into 30 categories. These include machine translation, semantics, information extraction, machine learning, parsing, sentiment analysis and opinion mining, resources and evaluation, natural language generation, dialogue, question answering, internet and social media NLP, summarization, representation learning and text representation, text mining, multimodal and multimedia NLP, information retrieval, discourse analysis, spoken language processing, NLP applications, syntax, grammar, word sense disambiguation, morphology, coreference resolution, word segmentation, phonetics, ethics and NLP, pragmatics, cognitive linguistics, and others. Based on this classification system, we analyze the differences in research difficulty among papers across different task categories.

We use the Kruskal–Wallis test (McKight & Najab, 2010). The results show significant differences in research difficulty among papers from different conferences at the $p < 0.05$ level. We select the top ten task categories by paper count and created box plots for research difficulty scores across these categories, as shown in Figure 3. The result shows that, between 2000 and 2021, machine learning papers in the NLP field show a wide range of research difficulty scores. It may due to the complexity of different research topics and methods and the technical depth and innovation required. After further analysis, we find that papers on machine learning and natural language generation have higher difficulty scores. These areas are harder than others. This is maybe because they need more complex methods and strong theories. Additionally, papers in machine translation and resources and evaluation tasks show more outliers, indicating that these research areas may present significant challenges and complexity.

**(3) Evolution analysis of research difficulty in academic papers within the NLP Field**

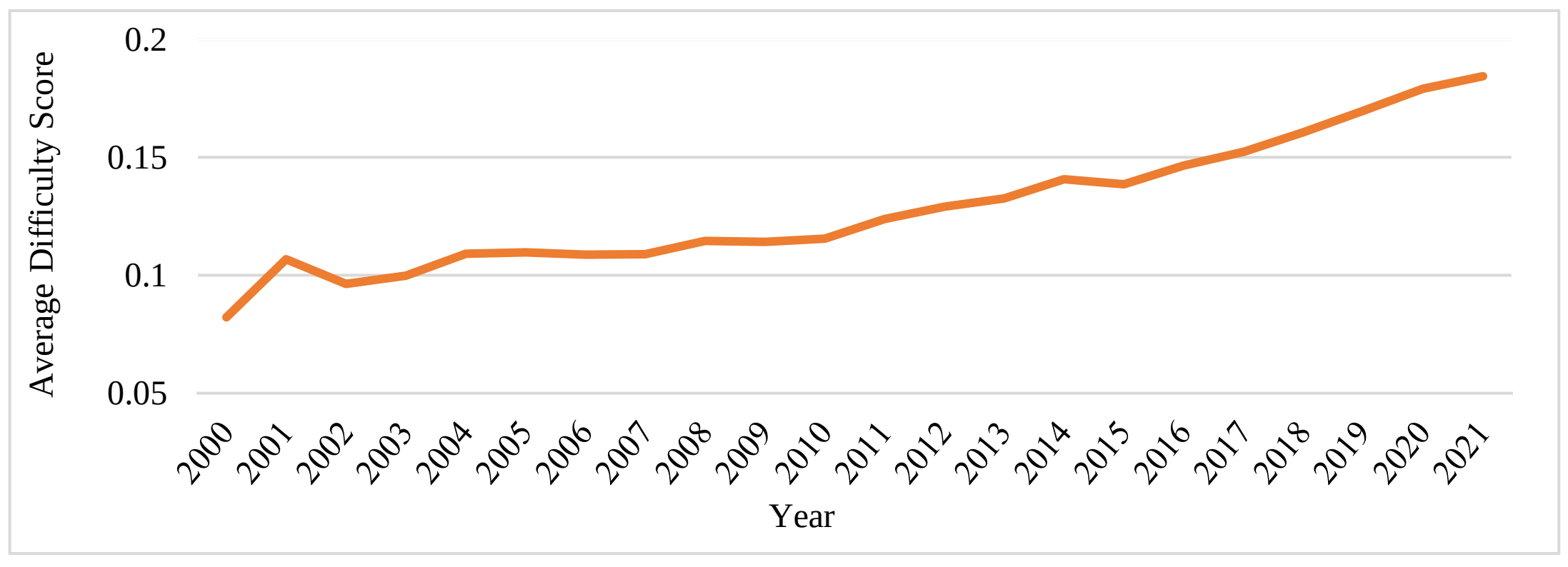

**Fig.4. The annual average research difficulty evolution diagram of academic papers in NLP field**

We calculate the average research difficulty score of academic papers in the NLP field by publication year, as shown in Figure 4. The annual average research difficulty score of academic papers shows an overall increasing trend over time.

First, as the NLP field developing, researchers may be more tend to explore related complex and deep problems. Second, continuous technological advancements, especially in deep learning and big data, have diversified and complicated research issues within NLP. Furthermore, as NLP intersects more with other fields, researchers need more diverse knowledge and skills. This makes research more difficult.

As NLP papers get more difficult to research, reseachers need to keep getting better at their work. And they should learn new technologies and theories to handle more difficult problems.

### 4.3 Analysis of the relationship between research difficulty and academic impact of academic papers

After including all variables mentioned in Section 3.4 in the regression, it is observed that all variables have a Variance Inflation Factor (VIF) value below 5. This indicates no severe multicollinearity issues. Thus, the selected independent and control variables are suitable for constructing the regression model. After analyzing, we note that the data on academic impact in the NLP field does not follow a normal distribution, with a relatively dispersed distribution and notable outliers. Some studies have explored the most appropriate citation count regression strategy. The results show that using a negative binomial regression model to explore the relationship between total citation frequency and various factors is not the best approach. Instead, a better strategy is to add 1 to the citation frequency, take its logarithm as the dependent variable, and then use a general linear (Ordinary Least Squares) model for regression (Bornmann et al., 2012). Therefore, we first apply a log transformation (adding 1) to the academic impact data and then uses a linear regression model to explore the relationship between research difficulty and academic impact.

Following Kuhn’s perspective that "normal science is the usual state of science, most scientists are normal scientists, and most of the work scientists do is normal science" (Kuhn, 1997), we initially hypothesize that research difficulty may have a positive impact on academic impact and may follow an inverted U-shaped curve relationship. Accordingly, after constructing a linear regression model (Model 1), we add the quadratic term of the independent variable as a predictor in a quadratic regression model (Model 2) to explore any potential curvilinear relationship between research difficulty and academic impact.

**Table 7. The results of regression analysis of research difficulty and academic impact of papers**

| Academic impact | Regression model（1） | Regression model（2） |
|---|---|---|
| Papers difficulty score | 3.206*** | 10.07*** |
| | (0.410) | (0.821) |
| The quadratic term of papers difficulty score | | -15.25*** |
| | | (1.686) |
| Page number of papers | 0.0780*** | 0.0596*** |
| | (0.008) | (0.008) |
| Number of references in papers | 0.0132*** | 0.0133*** |
| | (0.001) | (0.001) |
| Presence or absence of high-level institutions | 0.303*** | 0.303*** |
| | (0.030) | (0.030) |
| Modality for cooperation | | |
| CT1 | 0.190** | 0.139** |
| | (0.054) | (0.054) |
| CT2 | -0.0233 | -0.206*** |
| | (0.063) | (0.065) |
| CT3 | -0.415*** | -0.672*** |
| | (0.086) | (0.088) |
| Category of conference | | |
| EMNLP | -0.209*** | -0.211*** |
| | (0.029) | (0.028) |
| NAACL | -0.0477 | -0.0353 |
| | (0.034) | (0.034) |
| The fixed effect of publication year | YES | YES |

**Note**: *** indicates $p < 0.01$, ** indicates $p < 0.05$, * indicates $p < 0.1$. The decimals in parentheses represent robust standard errors. Regression coefficients are rounded to three significant figures, and robust standard errors are rounded to three decimal places.

We use the STATA regression analysis tool to build a linear regression model. Robust standard errors are incorporated into the regression model to better handle non-normally distributed data and provide reliable standard error estimates. The log-likelihood function is used to fit the data and estimate model parameters. In regression model (1), the standard error is 1.1147, the F-statistic is 63.78, and the corresponding p-value is less than 0.05. The goodness-of-fit R-squared value in regression model (1) is 0.1821. In regression model (2), the standard error is 1.1079, the F-statistic is 67.51, and the corresponding p-value is also less than 0.05. The R-squared value in regression model (2) is 0.1920. These results indicate that the independent variables in both regression models (1) and (2) significantly impact the dependent variable.

The results of the regression analysis are shown in Table 7. In regression model (1), the research difficulty score has a significant positive relationship with academic impact (linear term $r=3.206$, $p<0.01$). It indicates that increasing research difficulty may enhance a paper's academic impact. In regression model (2), the research difficulty score continues to show a positive effect on academic impact (linear term $r=10.07$, $p<0.01$), while the quadratic term has a significant negative correlation with academic impact (quadratic term $r=-15.25$, $p<0.01$). It suggests a potential inverted U-shaped relationship between the independent and dependent variables.

To further verify the presence of this inverted U-shaped relationship, we employed the test developed by Lind & Mehlum (2010), with results shown in Table 8. The test finds a lower-bound slope of 9.6827 and an upper-bound slope of -9.6747, confirming a pattern of initial increase followed by decline, with the t-test p-value below 0.01. This indicates a significant inverted U-shaped curve relationship between research difficulty and academic impact. In the early stages, higher research difficulty boosts academic impact. But once research difficulty goes beyond a certain point, increasing the difficulty of research will inhibit the academic impact of papers.

**Table 8. Test results of inverted U-shaped relationship between papers difficulty and academic impact of papers**

| Attribute | Lower limit | Upper limit | The overall test of inverted U-shaped relationship |
|---|---|---|---|
| Interval | 0.0127 | 0.6477 | |
| Slope | 9.6827 | -9.6747 | |
| t value | 3.0904 | -7.3411 | 7.34 |
| P>\|t\| | 1.59e-39 | 1.16e-13 | 1.16e-13 |
| | Extreme point： | 0.3304 | Existing an inverted U-shaped relationship |

Papers with low research difficulty usually cover basic or simple topics, so their contributions are limited. They are easy for many people to understand, but they might not have enough innovation or depth to make a big academic impact. That’s the reason they get fewer citations. Papers that are too difficult might use very complex theories or methods. Only a few experts can understand and use them, so they don’t get cited much by a wider audience. Papers with medium difficulty are just right. They’re challenging enough to be interesting, but not so hard that people can’t understand or use them. These papers often inspire more research and get cited more often. They balance theory and practice well, so their findings are useful and can be applied in different ways. This makes them more likely to be cited by other researchers and used in real-life situations, which increases their citation rates.

As shown in Table 7, paper length, the number of references, and involvement of high-ranking institutions are all significantly positively correlated with academic impact. This suggests that papers with more content, more references, and high-level institutional involvement are more likely to achieve higher academic impact. Papers involving high-ranking institutions tend to have greater academic influence as well.

Additionally, compared with research papers produced by a single institution, papers involving institutional collaboration can gain more influence, while papers involving international collaboration can only gain less influence. This shows that internal institutional collaboration is more conducive to enhancing the academic impact. Papers published at ACL tend to receive higher academic impact than those published at EMNLP. There are also substantial differences in academic impact across publication years, with papers published between 2016 and 2018 being more likely to achieve higher academic impact.

## 5. Conclusion and Future Works

This study aims to measure the research difficulty of academic papers and explore its relationship with academic impact. Using the NLP field as a case study, we constructed an evaluation system for assessing research difficulty in academic papers. Then we collected relevant data to calculate each paper's indicator values, and applied the entropy weight method to determine the weight of each indicator. Finally, we got a difficulty score for each paper. The findings show an upward trend in the research difficulty of NLP papers, with variations in difficulty across different task categories. Through regression analysis, the relationship between research difficulty and academic impact was examined. We invited experts in the NLP field to rate the research difficulty of sample papers conduct a correlation analysis with the measurement method used in this paper. Empirical results indicate that, in the NLP field, paper length, the number of references, and the involvement of high-ranking institutions are all significantly positively correlated with academic impact. The relationship between research difficulty and academic impact follows an inverted U-shaped curve. It suggests that increased research difficulty does not necessarily lead to higher academic impact.

This study has certain limitations. First, the constructed evaluation system may involve some subjectivity and may not fully reflect the actual research difficulty of papers. Future research will delve further into the concept of research difficulty and work toward building a more scientific and rational evaluation system. Second, our dataset includes only papers from the three major NLP conferences (ACL, EMNLP, and NAACL), making the scope relatively narrow. Future research could expand the dataset to cover more academic fields to more comprehensively explore the relationship between research difficulty and academic impact in academic papers. Third, this study does not explicitly distinguish between research difficulty and academic contribution in a theoretical sense. Future work will further explore their theoretical differentiation and modeling. Fourth, this study focuses on the NLP field, which has relatively clear disciplinary boundaries and limited interdisciplinary features. Future research will extend the proposed framework to representative interdisciplinary domains to further investigate and validate whether interdisciplinarity—particularly long-distance interdisciplinarity—constitutes a characteristic of research difficulty, both theoretically and empirically.